# Moore's Law is dead, long live Moore's Law！

Nick Zhang

Abstract

Moore's Law has been used by semiconductor industry as predicative indicators of the industry and it has become a self-fulfilling prophecy. Now more people tend to agree that the original Moore's Law started to falter. This paper proposes a possible quantitative modification to Moore's Law. It can cover other derivative laws of Moore's Law as well. It intends to more accurately predict the roadmap of chip's performance and energy consumption.

1. **A brief history of Moore's Law: Moore's Law is dead**

Moore's Law predicts that the number of transistors in integrated circuits doubles every fixed time period. Intuitively, exponential growth is not sustainable in long run. Although some economists have observed that sometimes the number of transistors increased even faster than Moore's Law, this is actually an illusion as some chips package multiple dies and die size have expanded slightly every ten years or so in the past. Transistor density in unit area, rather than the simple count of transistors in a chip, is a more meaningful measure.

In 1965, Moore predicted that the doubling time of transistor density would be one year [9], and ten years later in 1975, he revised it to two years [10], but now we have observed that the doubling time is more than three years. There are a few dominant factors in chip density, die size, linear dimension, technical cleverness (design), and technology innovations [15]. Moore's Law first appeared at the time of SSI (small-scale integration) when chips contained fewer than a hundred components. Moore's revision in mid 1970s happened in the period of MSI and LSI (medium- and large-scale integration) when chips contained fewer than 100,000 components. In SSI days, Moore's considerations were mainly economical rather than technical. As DRAM is the most important product in MSI and LSI period, and its structure is regular and repetitive. Production techniques, as major driving factor for density, were improved dramatically. Microprocessors are more sensitive to design as they are more complex.

Before 2000, the chip performance came from two sources. According to Denard Scaling, the frequency improvement of each generation of chips is about 40%, while the architecture advancement brought about by the transistor density of each generation conformed to Pollack Rule, that is, the square root level improvement, 41%. Each generation will basically double the performance as 1.4*1.41=1.97 [2]. Therefore, Moore's Law of transistor density can be converted into Moore's Law of performance. Denard Scaling has failed since 2005, and the progress of single-core performance cannot rely solely on higher frequency. Since then, multi-core processors have emerged to strengthen throughput performance. In short, the growth rate of transistor density is slowing down, and the frequency improvement is even harder, it is more difficult to increase performance. During the 20 years period before 2005, microprocessor performance advanced by 1,000 times, that is, it doubled every two years. However, it is obvious that this kind of exponential growth has not been maintained in the following ten years [13].

Moore's Law is not a mathematical theorem nor a law of physics. It is simply an observation. Its main significance is economic. Some say that it is a self-fulfilling prophecy of semiconductor industry. Its forecast power amazingly lasted for at least 20 years. Although Moore's predictions fit generally correct, his reasons were sometimes wrong, considering different factors played different roles in different periods.

Limits imposed by physics become prominent. Chip's performance improvement is mainly from advanced manufacturing process technology and new processor microarchitecture. As performance gets higher, so does power consumption. As Moore's Law deviates more and more from the earlier forecast, various derivative laws from Moore's Law, such as Koomey's Law [7], also failed to predict. In fact, the performance improvement is not a fixed proportion, but a variable. In this paper, a quantitative revision of Moore's Law is proposed so that the performance can be forecasted more accurately.

## 2. Revision of Moore's Law: Long live Moore's Law

If the ratio of annual improvement is constant, then the performance improvement with time follows an exponential function. Data show that the ratio of annual improvement decreases. This reflects that the improvement will continue, but it will become smaller and smaller. It is observed that the logarithm of annual improvement proportion can be fitted by $1/\ln(t)$, while the logarithm of the performance can be fitted by logarithmic integral, $li(t)$ ($\approx t/\ln(t)$). This revision of Moore's Law can explain the original version: when t is small, $t/\ln(t)$ is roughly a straight line, while when t is large, $t/\ln(t)$ will drift away from the straight line.

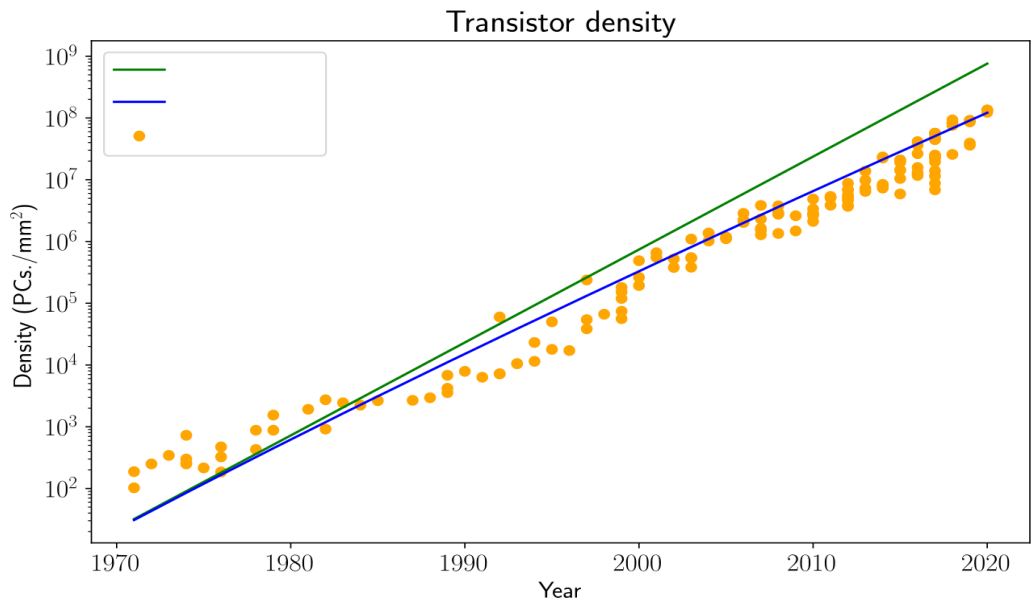

1. The increase of transistor density can be fitted by modified Moore's Law (blue).

Logarithmic integral function can be used to describe the phenomenon of "getting harder and harder". Transistor density can be fitted by the following formular:

$$y = a\frac{x'}{lnx'} + b,$$

where, $x' = x - 1943$

Koomey used to fit CPU energy efficiency data from the electron tube era. The constant here is just coincident that the ENIAC project started in 1943. It is interesting to note that Turing projected in 1950 [16] that by year 2000 the storage of a computer will reach capacity of $10^9$, i.e., gigabytes range. The modified Moore's Law, referred to as "Nick's Law"[14] for convenience, can also fit the chip performance data. To be fair, we firstly compare AMD and Intel's X86 architecture chips, to avoid factors brought by different architectures and manufacturing process technologies.

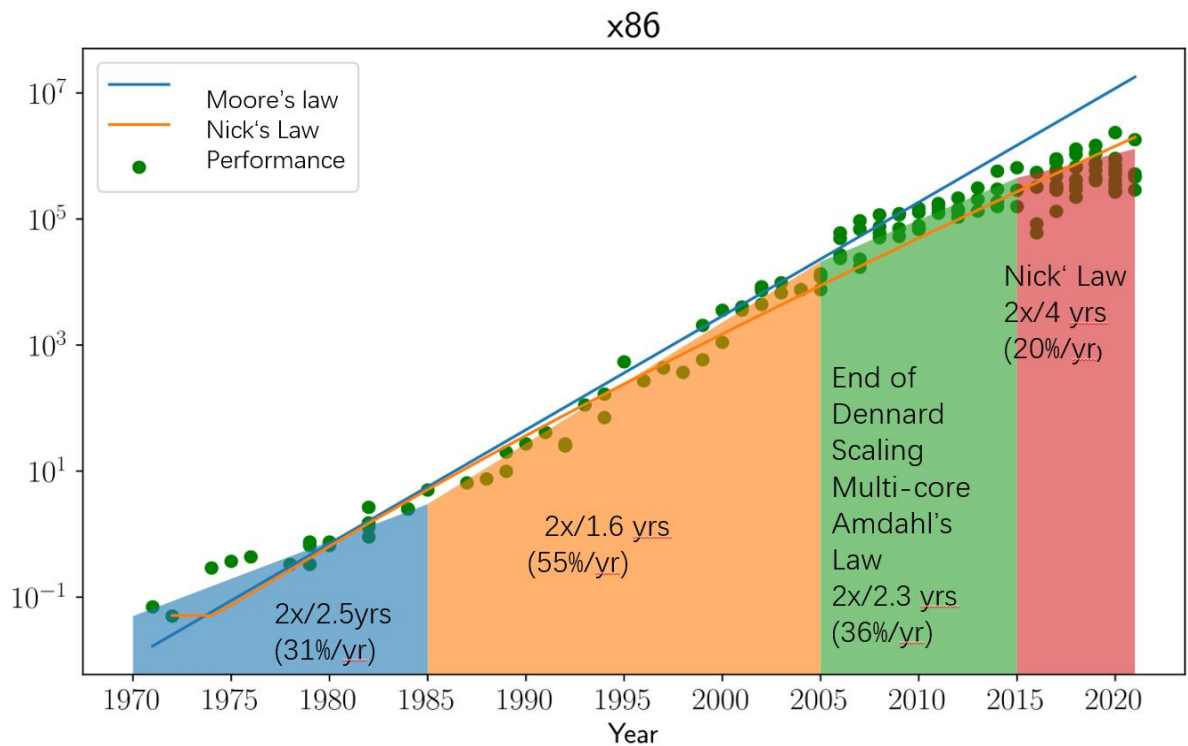

Figure 2. Modified Moore's Law can better fit CPU performance data.

Raw data of multi-core chips does not represent real performance. Firstly, not all problems can be parallelized. Amdahl's Law claims that the unparallel part is the bottleneck of parallel computing. Then, multi-core architecture needs parallel algorithms and corresponding software optimization. For primitive vector and matrix operations, it may have theoretical linear acceleration even though it is hard to achieve in practical scenario. However, most problems do not have linear speedup, nor even in theoretical sense. Parallel prefix sum is a classic textbook example, whose time complexity is O(log(n)), so the parallelism or speedup can reach O(n/log(n)), which is already surprisingly good. However, the typical parallel sorting algorithm, parallel merge-sort, can only achieve O(n/log$^2$(n)) [3].

In the past 15 years, the main frequency improvement of processors has been extremely slow. If we only look at the X86 architecture in the past 20 years, it can be fitted by log(t). SPECint benchmark measures single-core single-thread performance. It seems that log(t) can also fit logarithm of SPECint numbers.

The transistor density grows at a sub-exponential rate $O\left(2^{\frac{t}{\log t}}\right)$, and so does the performance. Before 2005, mainstream CPUs were almost all single cores, but now a wearable device easily packages eight cores. In the future, we will definitely see more cores, and require parallel algorithms and new programming paradigms. In order to achieve higher speedup, we must consider optimizing the processor architecture as a whole. Jim Keller's experience with X86 architecture is that every 3-5 years, microarchitecture should be redesigned. At present, memory is almost two orders of magnitude slower than processors. When the number of processors grows, data movement among different processors will make the problem worse. Various new DSA architectures are now designed to alleviate such problems. But this also leads to intricately multitude of software.

A recent study [17] on algorithmic improvement vs hardware improvement shows that nearly half of the algorithms have had no improvements, and only 14% of the algorithms have improved beyond Moore's Law.

### 3. Revision of Derivatives of Moore's Law's

With this modification to Moore's Law, derivative laws, such as Koomey's Law, shall be revised as well. The original Koomey's Law says that for the same computing power, the required energy consumption will be halved every certain time period, or conversely, computing power that can consume the same energy will be doubled every certain time period. Koomey's paper [7] in 2010, mainly used data from the economist Nordhaus[12]. The amount of data in that study was too small, and there were only about 20 data points in the 30 years period from 1980 to 2010. They compared the computing performance per kilowatt-hour of different supercomputers and PCs, and found that the difference was not that big, and the time needed to double the computing performance was basically between 1.52 and 1.57 years. In 2011, Koomey re-examined the historical data and revised the time interval to 2.6 years.

Recent data show that the reduction of energy consumption is becoming slower. In fact, certain architectures do have better energy efficiency than others. Nick's Law fits the latest data better. According to Landauer principle [1][8], logically irreversible computation must be accompanied by energy consumption, and the minimum energy consumption value is Landauer limit. According to the original Koomey's Law, Landauer limit will be reached by around 2050, then Koomey's Law will be completely invalid. This is precisely the time for EU and Japan to achieve carbon neutrality.

Koomey does not quite agree with restricting Koomey's Law with Landaurer limit. He believes that the only restriction of Koomey's law is people's cleverness rather than physics[6]. I find Moore's statement more interesting: "No exponential is forever: but 'forever' can be delayed!" [11]. Nothing can sustain the exponential growth, even human knowledge. This process of delay can be better fitted by Nick's law. Hence, the time to reach Landauer limit will be postponed further (see figure 3).

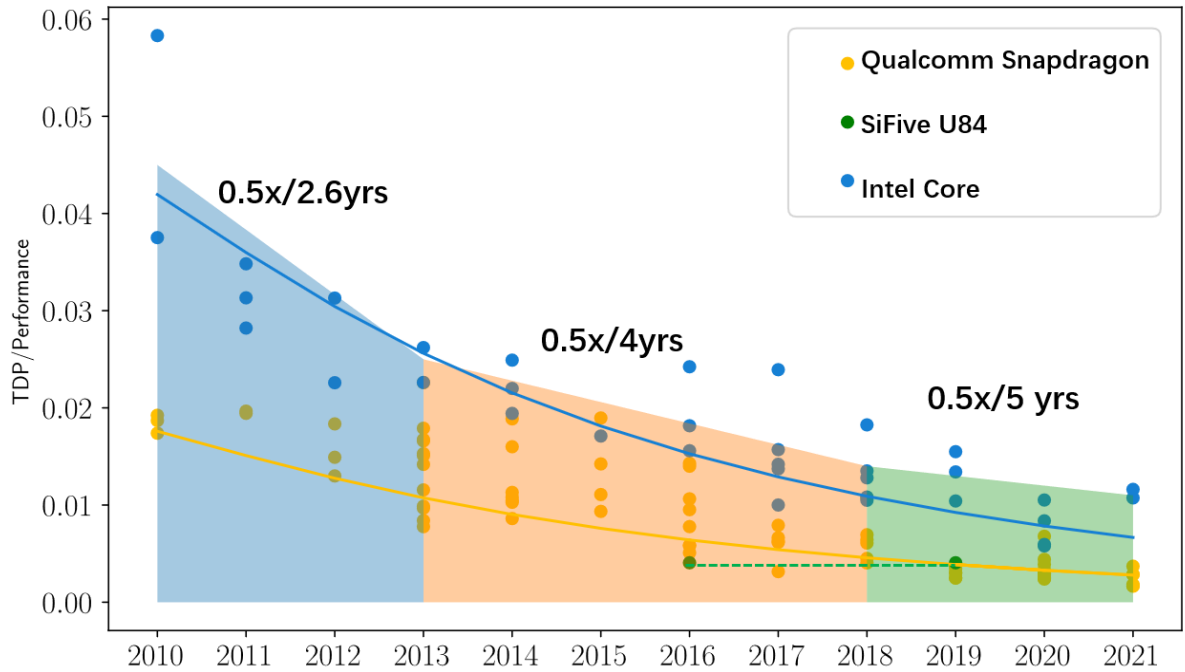

Figure 3. Modified Koomey's Law.

The sum of Top500 supercomputers' computational performance also deviates from Moore's Law more and more. Nick's law better fits the data.

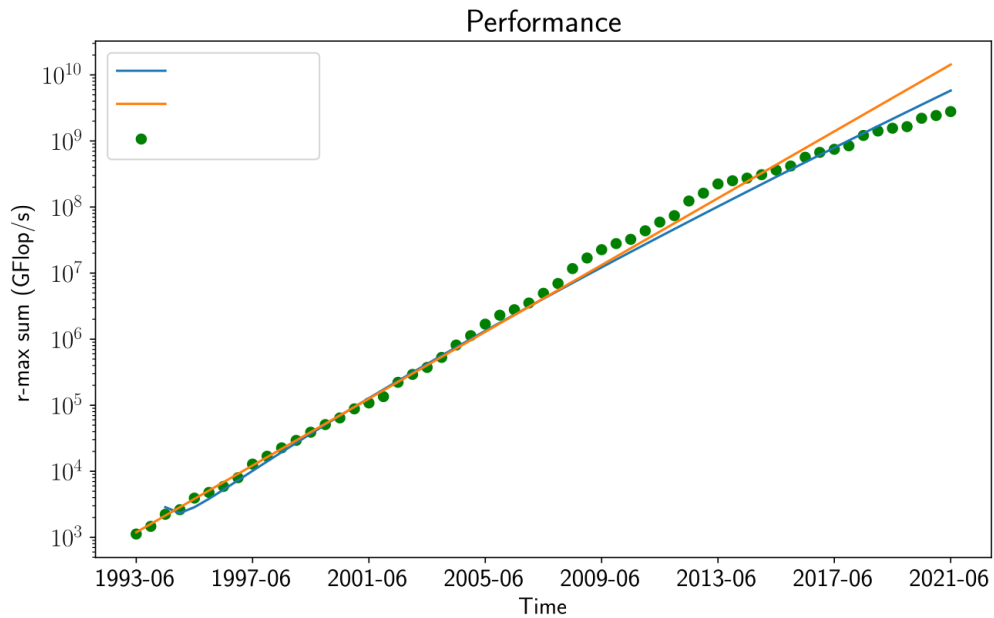

Fig. 4 Performance of Top500 supercomputers (the green dots are actual data, the orange curve is Moore's Law, the blue curve is modified fitting)

We can also apply Nick's law to energy efficiency of supercomputers, i.e., total computational performance/total energy consumption. At present, the energy consumption data of Top500 is not complete. We have to take data from Top10. The total energy consumption of supercomputers rises by log(t), and the power efficiency grows by $O\left(\frac{2^{\frac{t}{\log t}}}{\log t}\right)$. As data amount is small, the fitting curve is not as smooth.

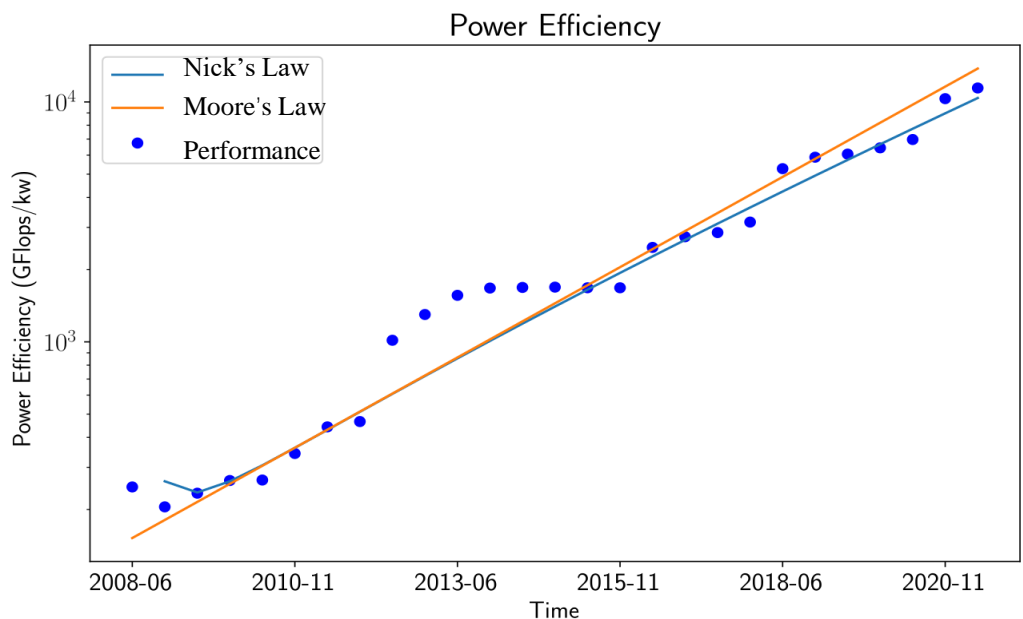

Figure 5. Power efficiency of supercomputers

Bell's Law[4] uses price to describe how different types of computing systems (referred to as *computer classes*) form, evolve and may eventually die out. At one particular period, there is an established class at roughly a constant price with increasing performance. Supercomputers, as another class, seek the best performance regardless of price. Bell estimate that every 10 years or so, a novel class with lower price will form and eventually replace the existing established classes. Bell accurately predicted that supercomputers would go into multi-computer era after 2000 [5]. In Cray's era, performance improvement was basically dependent upon clock frequency. While in the multi-computer era, performance comes from much more cheap CPU cores. ARM was regarded by Bell as a low-cost but novel architecture. Now ARM not only has a strong trend to replace the incumbent PC architecture, but also started to appear in supercomputers. The fastest supercomputer at present (2021), Fugaku, is powered by ARM.

Energy efficiency has now become a very important and urgent measure. Erik Brynjolfsson points out that Koomey's Law is more important than Moore's law. Bell's law forecasts that each class has a life cycle of 10-15 years. If we change Y- axis of Bell's Law to energy consumption, then the established class would be ARM, as mobile devices powered by ARM roughly maintain constant power consumption. The upper classes would be supercomputers and PCs, while the lower classes would be RISC-V and other novel architectures, with better energy efficiencies. Each class remains stable during its life cycle and then attempts to replace its upper class. The evolution is just like a paradigm shift or punctuated equilibrium.

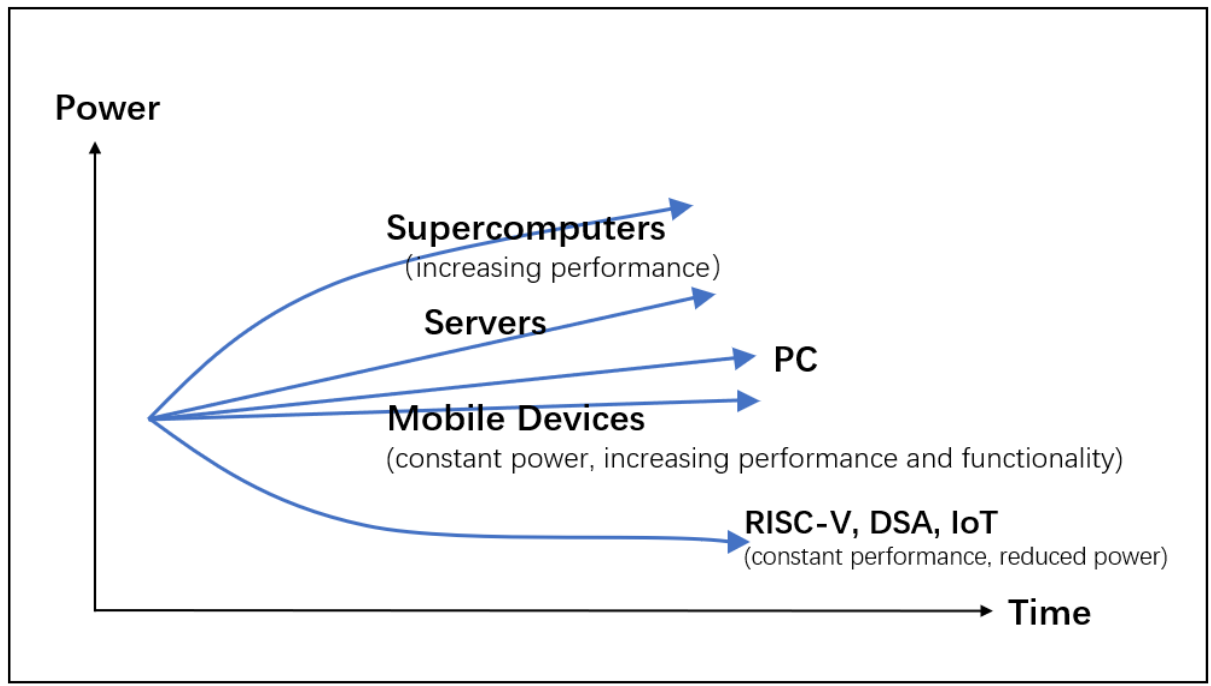

Fig. 6 Bell's Law based on power consumption (instead of price)

**Conclusion**

Moore's Law hits a wall. The wall is soft in the beginning but becomes harder and harder, like damping effect. The exponential improvement cannot last forever. Some people banter that "the number of people who believe Moore's Law is dead doubles every two years". In 2003, Moore pointed out that the doubling time of performance would be prolonged, but the growth rate of semiconductor industry still far exceeded that of almost all other industries [11]. The revised Moore's Law, or Nick's Law rather, attempts to state that the we can still be enjoying high growth, but the speed of it is gradually slowing down. We cautiously bet that the Nick's Law can last for another 20 years.